\documentclass[epj,draft]{svjour}{}
\pdfoutput=1
\usepackage{graphics}
\usepackage{amsmath,amssymb} 
\usepackage{epsfig} 
\usepackage{color}
\usepackage{bm} 
\usepackage{epstopdf}
\usepackage{rotating}
\usepackage{physics}
\usepackage{braket}
\usepackage{multirow}
\usepackage{hhline}
\usepackage{cite}
\usepackage{colortbl}
\usepackage[table]{xcolor}
\usepackage{booktabs}
\usepackage{graphicx}
\usepackage{slashed} 
\usepackage[british]{babel}
\usepackage[htt]{hyphenat}
 \usepackage[utf8]{inputenc}

\begin{document}

\title{Spin Partners of the $B^{(*)}\bar{B}^{(*)}$ resonances with a different approach than the Breit-Wigner parameterization}
\author{Duygu Yıldırım}
\institute{Physics Department, Faculty of Sciences and Arts, Amasya University, Amasya, Turkey\\
\email{ yildirimyilmaz@amasya.edu.tr }}
\titlerunning{Spin Partners of the $B ^{(*)}\bar{B}^{(*)}$ resonances with a different approach than the Breit-Wigner parameterization }
 \authorrunning{Duygu Yıldırım}
\abstract{ In general, resonances are obtained by the Breit-Wigner parameterization. However, it is not entirely appropriate to use this parameterization with near-threshold resonances such as the $Z_b(10610)$ and the $Z_b(10650)$, as Breit-Wigner does not contain the threshold effect. To eliminate this defect, a recently proposed alternative distribution, the Sill, is used to predict possible heavy quark spin symmetry partners of the $Z_b(10610)$ and the $Z_b(10650)$. With the Sill values of the $Z_b(10650)$ and the $Z_b(10610)$  states, assuming these exotics generate molecular states consisting of contact and a pion exchange potential, heavy quark spin symmetry partners are examined with $S$ and $D$-wave contributions. In the light of the Sill type resonance approach, all the partners are found as bound states. 
\\
}
\maketitle 

\section{\label{sec:level1}Introduction}

In the last two decades, numerous exotic hadron states have been observed. It has been rather challenging to put some of these states into classifications. A great deal of effort has been spent on especially the $\chi_{c1}(3872)$ in the charm sector and the $Z_b(10610)$ and  $Z_b(10650)$ in the bottom sector in this regard. The isovector $Z_b(10610)$ and $Z_b(10650)$ resonances are discovered by Belle Collaboration \cite{PhysRevLett.108.122001} in the invariant mass distribution of the $\Upsilon (5S)$ decays in $\pi^{\pm}h_b(mP)$ $(m=1,2,3)$ and $\pi^{\pm}\Upsilon(nS)$ $(n=1,2)$. These two bottomonium like states reside significantly close to open bottom  $B\bar{B}^*$($10604.6$ MeV) and $B^*\bar{B}^*$($10650.2$ MeV) threshold, respectively. Both states' quantum numbers are determined as $I^G(J^{PC})=(1^+)1^{+-}$ ~\cite{bellecollaboration2011observation}. What the $Z_b(10610)$ and $Z_b(10650)$ genuinely are has been questioned since their discovery. The existence of the observed heavy bottom states has been investigated with different techniques and assumptions. For instance, their proximity to the corres\-ponding thresholds have led the authors of Refs.~\cite{Bugg:2011jr, PhysRevD.91.034009} to consider them as the cusp effect. On the one hand, tetraquark assumptions of them have been studied ~\cite{Guo:2011gu,  Ali:2011ug,  Cui:2011fj,  Braaten:2014qka}. From the QCD sum rule perspective, a tetraquark state is found near $10500$ MeV ~\cite{Cui:2011fj}, with quantum numbers $J^{PC}=1^{+-}$. The region of the mass found is far below where the $Z_b(10610)$ and the $Z_b(10650)$(to be called $Z_b$ and $Z_b^{\prime}$) are; hence the tetraquark assumption is not entirely reliable. Also, they can not be bottomonium since they are elec\-trically charged and isovector states. On the other hand, because of being lying close to the $B^{(*)}\bar{B}^{(*)}$ threshold, the most assumed motivation for the $Z_b$ and the $Z_b^{\prime}$ are molecules. In addition, the quantum numbers are compa\-tible with the molecular assumption. The feature of decaying the $Z_b^{(\prime)}$ into the open flavor channel also supports the molecular idea~\cite{Guo:2017jvc}. Besides in the literature, several articles suggest the $Z_b^{(\prime)}$ are favored as molecular~\cite{Sun:2011uh,Wang:2014gwa,Dias:2014pva, Zhang:2011jja, Cleven:2013sq}. Even though the most plausible explanation seems that they are a molecule, the interpretation of the $Z_b^{(\prime)}$ as a molecular state is not apparent because experimental mass values for both states are above the related threshold of approximately $3$ MeV.

Many resonances have been searched with different resonance methods. The Breit-Wigner (BW) and the Flatt\'e para\-meterization and the effective-range expansion framework are some basic methods used to study heavy resonances. The generally accepted method is the Breit-Wigner para\-metrization, and experimental data are primarily analyzed with this approach \cite{PhysRevLett.108.122001, BaBar:2008qzi, ParticleDataGroup:2020ssz}. These methods differ from each other when it comes to threshold effects. While the Breit-Wigner and the Flatt\'e parameterization are affected by threshold effects, effective-range expansion is not affected by nearby thresholds.  The usage of the effective range expansion parameterization is more common than the Flatt\'e parametrization because the Flatt\'e parametrization can create negative effective ranges~\cite{Kang:2016jxw}. As for the effects of nearby threshold, Ref.~\cite{Kang:2016jxw} attached importance to this point and studied the $\chi_{c1}(3872)$, with the presence of near-threshold resonances, with effective-range expansion approximation up to and including the effective-range contribution. It was indicated for a more precise conclusion about the $\chi_{c1}(3872)$ that more data is needed.  By taking into account the thres\-hold effect, the authors of Ref.~\cite{Kang:2016ezb} studied the $Z_b^{(\prime)}$ states with effective-range expansion. They stated that the Breit-\-Wig\-ner function is in line with the data~\cite{Adachi:2011mks, PhysRevLett.108.122001} when the resonance is above the related threshold. Refs.~\cite{Cleven:2011gp, Cleven:2013rkf} suggested that if the $Z_b^{(\prime)}$ are considered a molecular state, the data should not be interpreted with the Breit-Wigner para\-metrization. In short, the Breit-Wigner para\-meteriza\-tion contains some intricacies. Recently, an alternative method called the Sill~\cite{Giacosa:2021mbz} was proposed for the near-threshold resonances instead of the Breit-Wigner. That work has shown that the Sill's approach is better in some circumstan\-ces, including XYZ states. 

For the first time, the possibility of a meson-meson bound state by exchanging with their light quarks was proposed by Voloshin and Okun~\cite{Voloshin:1976ap}. With the help of this idea, the effective field theory (EFT) of nuclear physics can portray the heavy meson systems. The interactions can be restrained by heavy quark spin symmetry (HQSS) thanks to heavy resonance content. The states concerned through this work are created by contact interaction, i.e., four meson vertices and one pion exchange. The power counting expansion in the EFT formalism indicates that contact interaction is the leading order, and one pion exchange (OPE) is the main subleading order correction.

This paper aims to investigate the effects of the Sill approach on the heavy quark spin partner predictions of the two exotic states. Throughout this paper, the $Z_b^{(\prime)}$ states, the $\chi_{c1}(3872)$  and the $\chi_{c0}(3915)$ are considered meson molecules. We reconsider these states with a different resonance dist\-ribution function and try to see the effects of OPE on the spin partners. With the help of the Sill, new predictions will be made for HQSS partners. It will be examined the impact on sub-order corrections to spin partners. The first two sub-order contributions are recalculated with the new values obtained by the Sill method. We primarily focused on the effect of estimated new resonance positions from the Sill approach and wave contributions of $S$ and $S$-$D$ on HQSS partners.
\section{\label{sec:level2} Formalism}

Resonance parameters, mass and decay width, rely on the description of resonance. It means that when the definition of resonance changes, the parameters change. The most applied and accepted description of resonance is the Breit-Wigner distribution formula. The known the Breit-Wigner formula  
\begin{equation}
    d^{BW}(E)=\frac{\Gamma}{2\pi} \frac{1}{(E-M)^2+\frac{\Gamma ^2}{4}} \, ,
\end{equation}
where $M$ is the mass of the resonance and $\Gamma$ the decay width.This formula is mainly used to define unstable states. Though the formula is very well applied to most resonances, particularly fundamental particles like $Z^0$ and $W^{\pm}$, it has some defects regarding near-threshold resonances and complex structures. For instance, the resonance area is disguised by the threshold effect; therefore, usage of the Breit-Wigner formalism differs from unitarity and analyticity. Furthermore, the branching ratios obtained by the Breit-Wigner parametrization for near-threshold states might not depict the decay probabilities~\cite{PhysRevD.93.034030}. It is stated in Ref.~\cite{Cleven:2011gp}, the Breit-Wigner formula should be resigned for the near-threshold states. Exploring which approach is more appropriate and applicable to a broader range is essential to understanding near-threshold exotic resonances.

A new formalism was put forward to solve the threshold problem that the Breit-Wigner essentially has. The distribution function was altered as in the Ref.~\cite{Giacosa:2021mbz}, which is called the Sill:
\begin{equation}
     d^{Sill}(E)= \frac{2E}{\pi} \frac{\sqrt{E^2-E^2_{th}}\tilde{\Gamma}}{(E^2-M^2)^2+(\sqrt{E^2-E^2_{th}}\tilde{\Gamma})^2}\theta(E-E_{th}) \, .
\end{equation}

The Sill holds threshold effects and can be applied to mesons and exotics because this new distribution is not relevant to the inner content of the hadron. The paper highlighted that when the Sill has applied to various ranges of resonance, it gives better results with experiments than the results obtained with the Breit-Wigner formalism. In addition, a method was proposed for the applicability of the Sill approach. The authors stated that, compared with the BW, the Sill is preferred when the ratio $\Delta m/ \Gamma$ is greater than $1/3$. The state's distance to the threshold is $\Delta m$, and the decay width of the state is $\Gamma$. This criterion also applies to cases the $Z_b^{(\prime)}$, the $\chi_{c1}(3872)$ and the $\chi_{c0}(3915)$. 

After the Sill approach is applied, any significant chan\-ges in pole positions of the $\chi_{c1}(3872)$, the $\chi_{c0}(3915)$, and the $X(3940)$ are not seen. On the other hand, as one can see in Table~\ref{tab:table1}, the changes in the positions of the $Z_b$ and $Z_b^{\prime}$ are striking. The Breit-Wigner values are approximately $3$ MeV above the respective thresholds, while the Sill masses of the $Z_b$ and $Z_b^{\prime}$ are $11$ and $3$ MeV below the thresholds, respectively. Therefore, the fact that they are pole positions below the threshold is quite suitable for them to be considered molecules. Molecular states, if the interaction does not rely on energy, generally reside below the threshold.

\begin{table}
\caption{\label{tab:table1} Mass and decay widths obtained by the Breit-Wigner parametrization's ~\cite{ParticleDataGroup:2020ssz} and the Sill's approach for the $Z_b(10610)$ and $Z_b(10650)$ resonances. Unit is MeV.}
\begin{tabular}{lll}
\firsthline
  & The BW ($E-i\Gamma /2 $)  & The Sill ($E-i\Gamma $) \\ \hline
$Z_b(10610)$  & $10607.2 - i18.4/2$    &  $10593.1 - i16.9$   \\
$Z_b(10650)$  &  $10652.2 - i11.5/2$   &  $10646.3 - i8.2$   \\
\lasthline
\end{tabular}
\end{table}
Except for the $Z_b$ and the $Z_b^{\prime}$, other heavy quark spin symmetry partners have negative $G$ parity and are labeled as $W_{bJ}$. These states are $W_{b0}$ and $W^{\prime}_{b0}$ for $J=0$, $W_{b1}$ for J=1, and  $W_{b2}$ for $J=2$.

The systems' symmetries shed light on how to solve the system's problems. The resonances in our focus contain heavy quarks whose masses $m_Q$ are much heavier than $\Lambda_{QCD}$. In the $m_Q \rightarrow \infty$ limit, the spin-related term in the Lagrangian of the system disappears. The symmetry brought by this circumstance provides that the total angular momentum of the heavy quark spin and the light degrees of freedom are conserved separately. The results of these two separate conversations indicate the existence of heavy quark spin partners with the same light cloud and different heavy quark spins. This symmetry in the literature is called heavy quark spin symmetry~\cite{manohar_wise_2000, NEUBERT1994259}. HQSS allows commenting that if one of the spin partners is observed experimentally, the others might be expected in the experiments.

The heavy quark mass limit makes heavy hadrons a static color source; therefore, interactions of heavy hadrons are flavor-independent. This symmetry, in the literature, is called heavy flavor symmetry (HFS). The noteworthy point here is that this symmetry makes it possible to slide between the charm and bottom sectors and use the same potential between meson interactions. 

For the heavy meson-antimeson states, the basis of various $J^{PC}$ values with partial waves are given below ~\cite{Nieves_2011,Nieves:2012tt}
\begin{eqnarray}\label{eq1}
    \mathcal{B}(0^{++}):&& P\bar{P}(^1 S_0), P^*\bar{P}^*(^1 S_0), P^*\bar{P}^*(^5 D_0)   \, ,
\end{eqnarray}
\begin{equation}\label{eq2} 
\begin{split}
     \mathcal{B}(1^{+-}): &P\bar{P}^*(^3 S_1,-), P\bar{P}^*(^3 D_1,-), P^* \bar{P}^*(^3 S_1), \\ & P^*\bar{P}^*(^3 D_1) \, ,
\end{split}
\end{equation}
\begin{equation}\label{eq3}
\mathcal{B}(1^{++}): P\bar{P}^*(^3 S_1,+), P\bar{P}^*(^3 D_1,+),\\  P^*\bar{P}^*(^5 D_1)    \, ,
\end{equation}
\begin{equation}\label{eq4}
\begin{aligned}
     \mathcal{B}(2^{++}):& P\bar{P}(^1 D_2), P\bar{P}^*(^3 D_2,+),   P^*\bar{P}^*(^5 S_2), \\ &P^*\bar{P}^*(^1 D_2), P^*\bar{P}^*(^5 D_2), P^*\bar{P}^*(^5 G_2)  \, ,
     \end{aligned}
\end{equation}
where the $P(\bar{P})$ and $P^*(\bar{P}^*)$  are heavy meson and heavy antimesons, respectively. The partial wave projections of the leading order potential are also found in Refs.~\cite{Nieves:2012tt, PhysRevD.85.114037}. For each partial wave, the $^{2S+1}L_J$ spectroscopic representation is employed; $S$, $L$, and $J$ are the intrinsic spin, the orbital angular momentum, and the total angular momentum of the two-meson system, respectively. For the considering isospin channel $I$, thanks to HQSS, there are two independent low energy constants (LEC): $C_{Ia}$ and $C_{Ib}$ \cite{AlFiky:2005jd}.

With power counting expansion, the contributions to the EFT potential can be sorted according to their importance. The heavy meson-antimeson potential expands as a power series of the ratio $Q/ \Lambda_0$
\begin{equation}
V=\sum_{\nu=0}^{\nu_{max}}V^{(\nu)}+\mathcal{O} \left[ \left( \frac{Q}{\Lambda_0}\right)^{\nu_{max}+1} \right] \, ,
\end{equation}
where $Q$ is the generic low energy scale, where the EFT is expected to perform,  and $\Lambda_0$ is the high energy scale.  The leading order short-range contribution is the contact interaction, while the long-range interaction at subleading order solely contains the OPE interaction. At the lowest order, which equates $\nu=0$,  in the given partial wave, the potentials in $J^{PC}$ sector are given \cite{manohar_wise_2000,Casalbuoni:1996pg,PhysRevD.68.054024,PhysRevD.45.R2188,PhysRevD.85.114037, Nieves:2012tt, AlFiky:2005jd}:
 \begin{equation}
V(0^{++})= \left(  \begin{array}{cc} C_{Ia}&\sqrt{3}C_{Ib} \\
 \sqrt{3}C_{Ib} & C_{Ia}-2C_{Ib}  \end{array}  \right) \, ,
 \end{equation}
 \begin{equation}
V(1^{+-})= \left(  \begin{array}{cc} C_{Ia}-C_{Ib}&2C_{Ib} \\
 2C_{Ib} & C_{Ia}-C_{Ib}  \end{array}  \right)  \, ,
 \end{equation}
 \begin{equation}
V(1^{++})= C_{Ia}+C_{Ib}  \, ,
 \end{equation}
 \begin{equation}
V(2^{++})= C_{Ia}+C_{Ib} \, .
\end{equation}
For the $1^{++}$ and $2^{++}$ sectors, the potentials of $P\bar{P}^*$ and $P^*\bar{P}^*$ states are equal. Considering that the coupled channel effects are next-to-next-to-leading order (N$^2$LO), for the $1^{+-}$ case, the $B\bar{B}^*$ and $B^*\bar{B}^*$ cases have identical LEC combinations. This equivalency states that the difference between the $B\bar{B}^*$ and $B^*\bar{B}^*$ thresholds should be the same as the difference between the isovector bottom states the $Z_b$ and $Z_b^{\prime}$, $46$ MeV. 

It should be noted that HQSS and HFS are approximate symmetries. Therefore, some deviation is expected from the heavy quark limit by an amount. The expected deviation order for LECs is $\Lambda _{QCD}/m_Q$. Hence, the error amount in the charm sector is around $20\%$ and in the bottom sector $7\%$. In short, HQSS is more reliable in the bottom sector than the charm sector. In other words, the large reduced mass of the $P^{(*)}\bar{P}^{(*)}$ means having more binding energy.

The kinetic energy of the heavy meson system is relatively small compared to its mass. In this case, the system is considered non-relativistic. The interaction potentials are iterated by the non-relativistic Lippmann-Schwinger equation to obtain the observables. The iterated potentials include contact and one pion exchange interactions in the given partial wave and with the given $J^{PC}$ sector. HQSS partners are searched by solutions of the Lippmann-Schwinger equation. The resonant pole positions reside where the inverse amplitude goes to zero the poles arise. The position of the pole is essential because if a pole locates below the threshold on the real axis of the first Riemann sheet, it is a bound state. If a pole is a virtual state, it appears below the threshold on the second Riemann sheet.

Long-range potential OPE and short-distance contact potential result in divergences in the Lippmann-Schwinger equation; therefore, it requires regularization. The problem is conducted by a Gaussian form factor $e^{-p^2 / \Lambda^2}$. Here, the cutoff $\Lambda$ basically has two assistance: it regulates the contact interaction and removes the singularity of OPE potential. The cutoff selection might influence the poles. For this reason, the choice of the cutoff is one of the keys to be considered. Phenomenologically, any cutoff value is acceptable. However, it should preserve the effects of the interested region's theoretical uncertainties at a given order~\cite{Epelbaum:2006pt}. For instance, the cutoff value should be chosen within the given range when considering the short-range effects.  In this work, the cutoff value is taken at about $1$ GeV. Besides the chosen value, all numerical results are examined with $0.5$ GeV and $1.5$ GeV cutoff values. It is seen that the results have a mild cutoff dependency. This verifies that the regularization process is successfully implemented.

One pion exchange potential of heavy meson channel can be written, in its most general form, as follows
\begin{equation}
   V(\vec{q})=\eta \frac{g^2}{2f_{\pi }^2} \vec{\tau}_1 \cdot \vec{\tau}_2 \frac{\vec{a}_1\cdot\vec{q}  \vec{b}_2\cdot \vec{q} }{q^2 +\mu ^2} ,
\end{equation}
where $\eta$ is the intrinsic C-parity of the meson-antimeson system, $\vec{a}_{1(2)} $ is the polarization operators, which depend on the initial and the final state of the meson system. $g$ is the axial coupling between meson and pion, $f_{\pi}$ is pion decay constant. $\vec{\tau}_{1(2)}$ is isospin operators on the heavy meson system $1(2)$ and $\mu$ is the reduced mass of the heavy meson pair in the given channel. One meson exchange potentials except pion are N$^2$LO order, and their calculations are beyond the scope of the present work. For more details, the reader is referred to Refs.~\cite{Ohkoda:2011vj, Sun:2011uh, Baru_2017}.

The isospin factor of the potential varies on the system's isospin. The isospin factor $\vec{\tau}_1 \cdot \vec{\tau}_2$ takes the values $-3$ and $1$ for total isospin $0$ and $1$, respectively. The sign of the OPE factor is influential in terms of the potential contribution and thus OPE's contribution to the states.

The coupling constant $g_D$ is approximately $0.59$ for the $D$ meson, defined by the  $D^* \rightarrow D \pi$ decay. The coupling constant $g$ for $B^*B\pi$ would be different from the one for the charm sector because of $1/m_Q$ corrections with the heavy quark mass $m_Q$ \cite{PhysRevD.49.2490}. Because of the kinematical constraint, it is impossible to fix the $g_B$ coupling from the experiment. $ g_B$ coupling varies in the literature from 0.37-0.63 \cite{Abada_2004,  PhysRevD.76.114503,  BECIREVIC2009231, Bernardoni_2015,  PhysRevC.83.025205}. Therefore the coupling selection is essential because its value changes OPE potential directly. However, the effect of the limit $g_B$ values in the literature on the heavy partners is also examined, and no more than a few MeV changes are observed in either version. For numeric calculations,  $g_D$ is taken as approximately 0.6 \cite{CLEO:2001sxb} for $g_D$ and $0.55$ for $g_B$. The pion decay constant $f_{\pi}$ is taken $132$ MeV. 

The single pion exchange potential between the meson $B$ and $B^*$ results in a mixture of different angular momentum, $L$, and $L\pm2$, in the tensor component. The net effect of OPE can be very significant when considering $S-D$ wave tensor transitions. It can also be quite attractive or repulsive, and this causes some changes in pole positions.
\subsection{General Remarks}

From the perspective of the EFT, in a heavy hadron molecule, each of the constituent heavy hadrons will be unable to see the internal structure of the other heavy hadron. Thus, in this work, the $Z_b^{(\prime)}$, the $\chi_{c1}(3872)$ and the $\chi_{c0}(3915)$ are con\-si\-dered as hadronic molecules generated by the contact and OPE interaction between the hadrons with preserving EFT. Within this study, the binding energies of the $\chi_{c1}(3872)$,  the $\chi_{c0}(3915)$ and  the new Sill $Z_b^{(\prime)}$ are used as input. From the theoretical perspective, the binding energy should be independent of any regularization scheme.  In addition, all divergences and energy dependencies are assumed to mainly absorb by the energy-dependent low-energy constants within the renormalization processes~\cite{Nieves_2011,Nieves:2012tt,Sun:2011uh}. This expresses that $C^{(0)}=C^{(0)}(\Lambda)$. With the acceptance, other divergencies or dependencies reside in the next leading orders. Then low energy constants, used as input, are fixed to reproduce the binding energy of the $\chi_{c1}(3872)$,  $\chi_{c0}(3915)$ and  $Z_b ^{(\prime)}$ masses. 

PDG values ~\cite{Workman:2022ynf} indicate that the $Z_b$ and $Z_b^{\prime}$ are above the threshold; thus, it is hard to link the resonances with a molecule assumption. On the other hand, as shown in Table~\ref{tab:table1}, the Sill resonance values locate below the corresponding thresholds. Locating these poles under the related thresholds is worthy information for using HQSS to make predictions about HQSS partners. Unfortunately, this is not enough for the bottom sector to continue because these two states, the $Z_b$ and $Z_b^{\prime}$, have the same LEC, $C_{1a}-C_{1b}$. We do not have other input to consider; therefore, extra input is needed to make any inference. As for the charm sector, there is enough input to make inferences~\cite{Workman:2022ynf}.

Plenty of research has been done about the $Z_b^{(\prime)}$ and their partners. Ref.~\cite{Ohkoda:2011vj} found that with pion exchange potential, the $1^{+-}$ and $1^{++}$ states are bound states with $7.7$ MeV and $16.7$ MeV binding energies, respectively. Also, in that study, the $2^{++}$ state is found as a resonance state. With the authors including the pion exchange contributions in their analysis, they concluded that the OPE potential predominates the positions of the state. In addition to Ref.~\cite{Ohkoda:2011vj}, it is stated in Ref.~\cite{Yang:2011rp} that the $Z_b^{(\prime)}$s and also the $W_{b0}^{\prime}$ state are bound with a few MeV binding energy.

Most of these works on the $Z_b^{(\prime)}$ show or suggest that $W_{b0}^{\prime}$ is a bound state~\cite{Yang_2012,Wang:2019gtg,Sun:2011uh,Ding:2020dio}. In light of these studies, the other data needed can now be selected as $W_{b0}^{\prime}$. It is chosen as a bound state with an average binding energy of $3$ MeV. The Sill $Z_b^{(\prime)}$ values are below the corresponding resonances. Within this study, it is assumed that the $\chi_{c1}(3872)$, $\chi_{c0}(3915)$ and the Sill $Z_b^{(\prime)}$ are bound states. While the $Z_b$ is below $11$ MeV below the threshold, the $Z_b^{\prime}$ is below the $3$ MeV below the threshold. Therefore there are two options to choose from as an input set. The $Z_b$ with binding energy $11$ MeV and the selected $W_{b0}^{\prime}$ are appointed as \textit{Version} $1$ while $Z_b^{\prime}$ with $3$ MeV binding energy and the selected $W_{b0}^{\prime}$ are appointed as \textit{Version} $2$.
\section{Results and Discussion}

As can be seen in Table~\ref{tab:table2},  the pole positions with the Sill's approach have undergone scarcely changes on the $\chi_{c1}(3872)$ and its heavy partners, since the Sill values turn out the same as the Breit-Wigner values in the charm sector. Apart from different input sets, no apparent difference is observed in the contact and $S$-wave.  But there are a few changes in the states with the $D$-wave contribution. At $1^{+-}$  $D^*\bar{D}^*$ and $2^{++}$ $D^*\bar{D}^*$ channel, $D$-wave contribution makes the state push down about $10$ MeV. Contrary to the $1^{+-}$ $D^*\bar{D}^*$ channel, the effect of the $D$-wave on the $0^{++}$ $D\bar{D}$ channel is repulsive and makes the state roughly $12$  MeV less bound.

\begin{table*}[hp!]
\caption{\label{tab:table2} Contact, $S$-wave, and $S$-$D$ wave contribution to the HQSS partners in the charm sector. The last two columns display two different input sets. In the Exp.(The Sill) column, the experiment results~\cite{Workman:2022ynf}(the Sill values) are taken as input. The poles are calculated for $\Lambda=1$ GeV. The masses are given in units of MeV.}
\renewcommand{\arraystretch}{1.2}
\begin{tabular}{|*{11}{c|}}
\hline \hline
&   &  & & \multicolumn{3}{c|}{\textbf{Exp.}} & \multicolumn{3}{|c|}{\textbf{The Sill }}  \\ \cline{5-10}
$I^G(J^{PC})$ &$D^{(*)}\bar{D}^{(*)}$& $C_{0X}$& & Contact & $S$ & $S$-$D$ & Contact & $S$ & $S$-$D$\\ \hline
$0^-(0^{++})$ & $D\bar{D}$ & $C_{0a}$& -- & $3713$ & $3713$ & $3725$ & $3713$ & $3712$ & $3725$ \\  \cline{5-10}
$0^-(1^{++})$ & $D^{*}\bar{D}$ &$C_{0a}+C_{0b}$ &$\chi_{c1}(3872)$& \multicolumn{3}{c|}{Input-$3871.65$} &  \multicolumn{3}{c|}{Input-$3871.65$}  \\\cline{5-10}
 $0^+(1^{+-})$ & $D^{*}\bar{D}$ & $C_{0a}-C_{0b} $&-- & $3821$ & $3825$ & $3822$ & $3822$ & $3821$ & $3822$  \\ \cline{5-10}
$0^-(0^{++})$ & $D^{*}\bar{D}^{*}$ & $C_{0a}-2C_{0b}$&$\chi_{c0}(3915)$& \multicolumn{3}{c|}{ Input-$3921.7$} & \multicolumn{3}{c|}{ Input-$3922.2$}	\\ \cline{5-10}
$0^+(1^{+-})$ & $D^{*}\bar{D}^{*}$ & $C_{0a}-C_{0b}$ &$X(3940)$& $3958$ & $3957$ & $3948$  & $3959$ & $3958$ & $3948$ \\  \cline{5-10}
$0^-(2^{++})$ & $D^{*}\bar{D}^{*}$ & $C_{0a}+C_{0b}$ &--& $4012$ & $4011$ & $4001$ & $4012$ & $4011$ & $4001$
\\ \hline \hline
\end{tabular}
\end{table*}

Due to our changing pole positions with the Sill's approach in the bottom sector,  it has undergone promising changes on the $Z_b^{(\prime)}$ and their heavy partners. The approaches of the bottom states are divided into two groups, and the results are given in Table~\ref{tab:table3}. In both versions, all heavy quark spin partners are observed as bound states.  It has been observed that the effects of the S and S-D waves on the channels show diversity. For the \textit{Verison} $1$ in the case of the $1^{++}$, the state has moved away from the threshold of about $20$ MeV by the $S$-wave effect, making it a tightly bound state. On the contrary, in the other states except for the $1^{++}$ sector, the contribution of the $S$-wave on the partners is repulsive, bringing the poles closer to the corresponding thresholds.  While the effect of the $D$-wave is repulsive for $0^{++}$ channel, it remains a bound state; in the $1^{++}$ sector; in addition to $S$-wave contribution, the $D$-wave also provided an additional attractive effect, making the state deeply bound. With the addition of the total OPE interaction, the state locates $23$ MeV away from the threshold. At the $1^{+-}$ and $2^{++}$ $B^*\bar{B}^*$ channels, the $S$-wave and the $D$-wave appear to have canceled each other contributions. 

In \textit{Version} $2$, no significant changes are seen in the contact and the $S$-wave contribution, but meanwhile, two resonances are observed to change with the $D$-wave component. The $1^{++}$ $B\bar{B}^*$ and $1^{+-}$ $B\bar{B}^*$ channels became more tightly states with $10$ and $6$ MeV, respectively. It has been found that the masses of the partners in  \textit{Version} $2$ are generally bigger than  \textit{Version} $1$.  Our results with OPE in \textit{Version} 2 are rather close to the threshold. This situation is in harmony with the idea that they should be considered as hadronic molecules in the first place.  

In Ref.~\cite{Baru_2017}, the effect of HQSS breaking scale $\delta$ on the position of HQSS partners of the $Z_b^{(\prime)}$ is examined mainly. The paper states that the $D$-wave contribution could not be ignored. Besides, if just the $S$-wave considered reproducing the partners, the states wouldn't change significantly. In general, it is not expected that an isovector meson-antimeson is a bound state because of isospin contribution. In that paper, the $Z_b^{(\prime)}$' binding energies are taken as input, and except for the $W_{b0}^{\prime}$ state, the $W_{b0}$,  $W_{b1}$, and $W_{b2}$ have been found as bound states. In particular, it was stated that the $W_{b2}$ is a promising state that can be seen in experiments.  

\begin{table*}[hp!]
\caption{\label{tab:table3} Contact, $S$-wave, and $S$-$D$ wave contribution to the HQSS partners in the bottom sector. Last two columns display two different approaches; Version $1$ and Version $2$. The poles are calculated for $\Lambda=1$ GeV. The masses are given in units of MeV.}
\renewcommand{\arraystretch}{1.2}
\begin{tabular}{|*{11}{c|}}
\hline \hline
&   &  &  & \multicolumn{3}{c|}{\textbf{ Version} $1$} & \multicolumn{3}{|c|}{\textbf{ Version} $2$}  \\ \cline{5-10}
$I^G(J^{PC})$ &$B^{(*)}\bar{B}^{(*)}$& $C_{1X}$&  & Contact & $S$ & $S-D$ & Contact & $S$ & $S-D$ \\ \hline
$1^-(0^{++})$ & $B\bar{B}$ & $C_{1a}$ & $W_{b0}$ & $10537$ & $10541$ & $10544$ & $10556$ & $10556$ & $10559$ \\    \cline{5-10}
$1^-(1^{++})$ & $B^{*}\bar{B}$ &$C_{1a}+C_{1b}$ &$W_{b1}$ & $10596$ & $10576$ & $10573$  & $10601$ & $10601$ & $10591$  \\  \cline{5-10}
 $1^+(1^{+-})$ & $B^{*}\bar{B}$ & $C_{1a}-C_{1b} $&  $Z_{b}$ & \multicolumn{3}{c|}{ Input-$10593.1$} & $10601$ & $10600$ & $10594$  \\ \cline{5-10}
$1^-(0^{++})$ & $B^{*}\bar{B}^{*}$ & $C_{1a}-2C_{1b}$& $W_{b0}^{\prime}$ & \multicolumn{3}{c|}{ Input-$10646.4$} & \multicolumn{3}{c|}{ Input-$10646.4$}	\\   \cline{5-10}
$1^+(1^{+-})$ & $B^{*}\bar{B}^{*}$ & $C_{1a}-C_{1b}$ & $Z_{b}^{\prime}$ & $10638$ & $10640$ & $10634$  & \multicolumn{3}{c|}{ Input-$10646.3$} \\ \cline{5-10}
$1^-(2^{++})$ & $B^{*}\bar{B}^{*}$ & $C_{1a}+C_{1b}$ & $W_{b2}$& $10614$ & $10621$ & $10614$ & $10646$ & $10646$ & $10645$  \\ \hline \hline
\end{tabular}
\end{table*}

Within the HQSS perspective, the states should be more bound in the bottom sectors than the charm sectors. But the Sill results do not satisfy this expectation. On the other side, the mass difference between the Sill $Z_b$ and $Z_b^{\prime}$ is about $46$ MeV, as expected. This consistency increases the credibility of the prediction made here. 

In Refs.~\cite{Baru_2017,Bondar:2011ev,Cleven:2011gp}, it is assumed that the $Z_b^{(\prime)}$ are bound states, also in Ref.~\cite{Sun:2011uh} suggested that the isovector $B\bar{B}^*$ and $B^*\bar{B}^*$ states could exist as a $B\bar{B}^*$ molecular state. These works align with the result given here, all bound states. It is presented in the Ref.~\cite{Sun:2011uh} that the $B^*\bar{B}^*$ $2^{++}$ state is not likely to exist within reasonable cutoff values. Contrary to the results of Ref.~\cite{Baru_2017}, the state is also found deeply bound at $\Lambda=1$ GeV here. All in all, by employing the Sill approach, we might see the heavy partners in the experiments.


\clearpage

\begin{thebibliography}{100}

\bibitem{PhysRevLett.108.122001}
A.~Bondar~\textit{ et al.}
\newblock {Observation of two charged bottomoniumlike resonances in
  $\ensuremath{\Upsilon}(5s)$ decays}.
\newblock {\em Phys. Rev. Lett.}, 108:122001, Mar 2012.

\bibitem{bellecollaboration2011observation}
Belle Collaboration and I.~Adachi.
\newblock{Observation of two charged bottomonium-like resonances, 2011}.

\bibitem{Bugg:2011jr}
D.~V. Bugg.
\newblock {An Explanation of Belle states $Z_b(10610)$ and $Z_b(10650)$}.
\newblock {\em EPL}, 96(1):11002, 2011.

\bibitem{PhysRevD.91.034009}
E.~S.Swanson
\newblock {${Z}_{b}$ and ${Z}_{c}$ exotic states as coupled channel cusps}
\newblock {\em Phys. Rev. D}, 91(3):034009, 2015.

\bibitem{Guo:2011gu}
T.~ Guo, L.~Cao, M.~ Zhou, and H.~ Chen.
\newblock {The Possible candidates of tetraquark : $Z_b(10610)$ and $Z_b(10650)$}.
\newblock {6 2011}.

\bibitem{Ali:2011ug}
A.~ Ali, C.~ Hambrock, and W.~ Wang.
\newblock {Tetraquark Interpretation of the Charged Bottomonium-like states $Z_b^{\pm}(10610)$ and $Z_b^{\pm}(10650)$ and Implications}.
\newblock {\em Phys. Rev. D}, 85:054011, 2012.

\bibitem{Cui:2011fj}
C.-Y.~Cui, Y.-L.~ Liu, and M.-Q.~ Huang.
\newblock {Investigating different structures of the $Z_{b}$(10610) and $Z_{b}$(10650)}.
\newblock {\em Phys. Rev. D}, 85:074014, 2012.

\bibitem{Braaten:2014qka}
E.~ Braaten, C.~ Langmack, and D.~Hudson Smith.
\newblock {Born-Oppenheimer Approximation for the XYZ Mesons}.
\newblock {\em Phys. Rev. D}, 90(1):014044, 2014.

\bibitem{Guo:2017jvc}
F. -K.~ Guo, C.~ Hanhart, U.-G.  Mei\ss{}ner, Q.~ Wang, Q. Zhao, and
  B.-S.~ Zou.
\newblock {Hadronic molecules}.
\newblock {\em Rev. Mod. Phys.}, 90(1):015004, 2018.

\bibitem{Sun:2011uh}
Z.-F.~ Sun, J.~ He, X.~ Liu, Z.-G.~ Luo, and S.-L.~ Zhu.
\newblock {$Z_b(10610)^\pm$ and $Z_b(10650)^\pm$ as the $B^*\bar{B}$ and
  $B^*\bar{B}^{*}$ molecular states}.
\newblock {\em Phys. Rev. D}, 84:054002, 2011.

\bibitem{Wang:2014gwa}
Z.-G.~ Wang.
\newblock {Reanalysis of the $Y(3940)$, $Y(4140)$, $Z_c(4020)$, $Z_c(4025)$ and
  $Z_b(10650)$ as molecular states with QCD sum rules}.
\newblock {\em Eur. Phys. J. C}, 74(7):2963, 2014.

\bibitem{Dias:2014pva}
J.~M. Dias, F.~Aceti, and E.~Oset.
\newblock {Study of $B\bar{B}^*$ and $B^*\bar{B}^*$ interactions in $I=1$ and
  relationship to the $Z_b(10610)$, $Z_b(10650)$ states}.
\newblock {\em Phys. Rev. D}, 91(7):076001, 2015.

\bibitem{Zhang:2011jja}
J.-R.~ Zhang, M.~ Zhong, and M.-Q.~ Huang.
\newblock {Could $Z_{b}(10610)$ be a $B^{*}\bar{B}$ molecular state?}
\newblock {\em Phys. Lett. B}, 704:312--315, 2011.

\bibitem{Cleven:2013sq}
M.~ Cleven, Q.~ Wang, F.-K.~ Guo, C.~ Hanhart, U.-G. ~Meissner, and
  Q.~ Zhao.
\newblock {Confirming the molecular nature of the $Z_b(10610)$ and the
  $Z_b(10650)$}.
\newblock {\em Phys. Rev. D}, 87(7):074006, 2013.

\bibitem{BaBar:2008qzi}
B.~ Aubert et~al.
\newblock {A Study of $B \to X(3872) K$, with $X(3872) \to J/\Psi \pi^{+}
  \pi^{-}$}.
\newblock {\em Phys. Rev. D}, 77:111101, 2008.

\bibitem{ParticleDataGroup:2020ssz}
P.~A. Zyla et~al.
\newblock {Review of Particle Physics}.
\newblock {\em PTEP}, 2020(8):083C01, 2020.

\bibitem{Kang:2016jxw}
Xian-Wei Kang and J.~A. Oller.
\newblock {Different pole structures in line shapes of the $X(3872)$}.
\newblock {\em Eur. Phys. J. C}, 77(6):399, 2017.

\bibitem{Kang:2016ezb}
X.-W.~ Kang, Z.-H.~ Guo, and J.~A. Oller.
\newblock {General considerations on the nature of $Z_b(10610)$ and
  $Z_b(10650)$ from their pole positions}.
\newblock {\em Phys. Rev. D}, 94(1):014012, 2016.

\bibitem{Adachi:2011mks}
I.~Adachi.
\newblock {Observation of two charged bottomonium-like resonances}.
\newblock {\em {9th Conference on Flavor Physics and CP Violation}}, 5 2011.

\bibitem{Cleven:2011gp}
M.~ Cleven, F.-K.~ Guo, C.~ Hanhart, and U.-G. ~Meissner.
\newblock {Bound state nature of the exotic $Z_b$ states}.
\newblock {\em Eur. Phys. J. A}, 47:120, 2011.

\bibitem{Cleven:2013rkf}
M.~ Cleven.
\newblock {\em {Systematic Study of Hadronic Molecules in the Heavy-Quark
  Sector}}.
\newblock {PhD thesis, Bonn U., 2013}.

\bibitem{Giacosa:2021mbz}
F.~ Giacosa, A.~ Okopi\'nska, and V.~ Shastry.
\newblock {A simple alternative to the relativistic Breit\textendash{}Wigner
  distribution}.
\newblock {\em Eur. Phys. J. A}, 57(12):336, 2021.

\bibitem{Voloshin:1976ap}
M.~B. Voloshin and L.~B. Okun.
\newblock {Hadron Molecules and Charmonium Atom}.
\newblock {\em JETP Lett.}, 23:333--336, 1976.

\bibitem{PhysRevD.93.034030}
Y.H.~ Chen, J.T. ~Daub, F.-K.~ Guo, B. Kubis, U.-G. ~Mei\ss{}ner,
  and Bing-Song Zou.
\newblock {Effect of ${Z}_{b}$ states on
  $\mathrm{\ensuremath{\Upsilon}}(3s)\ensuremath{\rightarrow}\mathrm{\ensuremath{\Upsilon}}(1s)\ensuremath{\pi}\ensuremath{\pi}$ decays}.
\newblock {\em Phys. Rev. D}, 93:034030, Feb 2016.

\bibitem{manohar_wise_2000}
A.V. ~Manohar and M.B. ~Wise.
\newblock {\em Heavy Quark Physics}.
\newblock {Cambridge Monographs on Particle Physics, Nuclear Physics and Cosmology. Cambridge University Press, 2000}.

\bibitem{NEUBERT1994259}
M.~ Neubert.
\newblock {Heavy-quark symmetry}.
\newblock {\em Physics Reports}, 245(5):259--395, 1994.  

\bibitem{Nieves_2011}
J.~Nieves and M.P.~Valderrama.
\newblock { The existence of $B\bar{B}^*$ bound states from the
  $X(3872)$ and heavy quark symmetry}.
\newblock {\em Physical Review D}, 84(5), Sep 2011.

\bibitem{Nieves:2012tt}
J.~Nieves and M. P.~ Valderrama.
\newblock {The Heavy Quark Spin Symmetry Partners of the X(3872)}.
\newblock {\em Phys. Rev. D}, 86:056004, 2012.

\bibitem{PhysRevD.85.114037}
M.P.~Valderrama.
\newblock {Power counting and perturbative one pion exchange in heavy meson molecules}.
\newblock {\em Phys. Rev. D}, 85:114037, Jun 2012.

\bibitem{AlFiky:2005jd}
M.T.~ AlFiky, F.~ Gabbiani, and A.A.~ Petrov.
\newblock {$X(3872):$ Hadronic molecules in effective field theory}.
\newblock {\em Phys. Lett. B}, 640:238--245, 2006.

\bibitem{Casalbuoni:1996pg}
R.~Casalbuoni, A.~Deandrea, N.~Di~Bartolomeo, Raoul Gatto, F.~Feruglio, and
  G.~Nardulli.
\newblock {Phenomenology of heavy meson chiral Lagrangians}.
\newblock {\em Phys. Rept.}, 281:145--238, 1997.

\bibitem{PhysRevD.68.054024}
W. A.~ Bardeen, E.J. ~Eichten, and C. T. ~Hill.
\newblock {Chiral multiplets of heavy-light mesons}.
\newblock {\em Phys. Rev. D}, 68:054024, Sep 2003.

\bibitem{PhysRevD.45.R2188}
M.-B. ~Wise.
\newblock {Chiral perturbation theory for hadrons containing a heavy quark}.
\newblock {\em Phys. Rev. D}, 45:R2188--R2191, Apr 1992.

\bibitem{Ohkoda:2011vj}
S.~ Ohkoda, Y.~ Yamaguchi, S.~ Yasui, K.~ Sudoh, and A. Hosaka.
\newblock {Exotic mesons with hidden bottom near thresholds}.
\newblock {\em Phys. Rev. D}, 86:014004, 2012.


\bibitem{Baru_2017}
V.~Baru, E.~Epelbaum, A.A. ~Filin, C.~Hanhart, and A.V.~ Nefediev.
\newblock {Spin partners of the $\ensuremath{Z_b(10610)}$ and
  $\ensuremath{Z_b(10650)}$ revisited}.
\newblock {\em Journal of High Energy Physics}, 2017(6), Jun 2017.

\bibitem{PhysRevD.49.2490}
H.-Y.  ~Cheng, C.-Y.~ Cheung, G.-L.~ Lin, Y.~C. Lin, T.-M. ~ Yan, and
  H.-L.~ Yu.
\newblock {Corrections to chiral dynamics of heavy hadrons: $\frac{1}{M}$
  correction}.
\newblock {\em Phys. Rev. D}, 49:2490--2507, Mar 1994.

\bibitem{Abada_2004}
A.~Abada, D.~Beirevi, P.~Boucaud, G.~Herdoiza, J.P Leroy, A.L.~ Yaouanc, and O.~Pene.
\newblock {Lattice measurement of the couplings $g_{\infty}$ and $g_{D^*D \pi}$}.
\newblock {\em Journal of High Energy Physics}, 2004(02):016, feb 2004.

\bibitem{PhysRevD.76.114503}
W.~ Detmold, K.~ Orginos, and M.-J.~ Savage.
\newblock {$bb$ potentials in quenched lattice qcd}.
\newblock {\em Phys. Rev. D}, 76:114503, Dec 2007.

\bibitem{BECIREVIC2009231}
D.~ Bećirević, B.~ Blossier, E.~ Chang, and B.~ Haas.
\newblock  {$g_{B^{*}B \pi}$-coupling in the static heavy quark limit.} 
\newblock {\em Physics Letters B}, 679(3):231--236, 2009.

\bibitem{Bernardoni_2015}
F.~ Bernardoni, J.~ Bulava, M.~ Donnellan, and R.~ Sommer.
\newblock {Precision lattice qcd computation of the $b^{⁎} b \pi$ coupling}.
\newblock {\em Physics Letters B}, 740:278–284, Jan 2015.

\bibitem{PhysRevC.83.025205}
B. ~El-Bennich, M. A.~ Ivanov, and C. D.~ Roberts.
\newblock {Strong ${D}^{*}\ensuremath{\rightarrow} d \ensuremath{\pi}$ and
  ${B}^{*}\ensuremath{\rightarrow} b \ensuremath{\pi}$ couplings}.
\newblock {\em Phys. Rev. C}, 83:025205, Feb 2011.

\bibitem{CLEO:2001sxb}
A.~Anastassov et~al.
\newblock {First measurement of $\Gamma$($D^{*+}$) and precision measurement of
  $m_{D^{*+}} - m_{D^0}$}.
\newblock {\em Phys. Rev. D}, 65:032003, 2002.

\bibitem{Workman:2022ynf}
R.~L. Workman  et al.
\newblock {Review of Particle Physics}.
\newblock {\em PTEP}, 2022:083C01, 2022.

\bibitem{Yang:2011rp}
Y.-C.~ Yang, J.~ Ping, C.~ Deng, and H.-S.~ Zong.
\newblock {Possible interpretation of the $\ensuremath{Z_b(10610)}$ and
  $\ensuremath{Z_b(10650)}$ in a chiral quark model}.
\newblock {\em J. Phys. G}, 39:105001, 2012.

\bibitem{Yang_2012}
Y.-C. ~Yang, J..~ Ping, C.~ Deng, and H.-S.~ Zong.
\newblock {Possible interpretation of the $z_b(10610)$ and $z_b(10650)$ in a
  chiral quark model}.
\newblock {\em Journal of Physics G: Nuclear and Particle Physics},
  39(10):105001, Aug 2012.

\bibitem{Wang:2019gtg}
Q.~Wang, V.~Baru, E.~Epelbaum, A.~A. Fillin, C.~Hanhart, A.~V. Nefediev, and
  J.~L. Wynen.
\newblock {Implications of spin symmetry for $XYZ$ states}.
\newblock In {\em {18th International Conference on Hadron Spectroscopy and
  Structure}}, pages 254--258, 2020.

\bibitem{Ding:2020dio}
Z.-M.~ Ding, H.-Y.~ Jiang, and J.~ He.
\newblock {Molecular states from $D^{(*)}\bar{D}^{(*)}/B^{(*)}\bar{B}^{(*)}$
  and $D^{(*)}D^{(*)}/\bar{B}^{(*)}\bar{B}^{(*)}$ interactions}.
\newblock {\em Eur. Phys. J. C}, 80(12):1179, 2020.

\bibitem{Bondar:2011ev}
A.~E. Bondar, A.~Garmash, A.~I. Milstein, R.~Mizuk, and M.~B. Voloshin.
\newblock {Heavy quark spin structure in $Z_b$ resonances}.
\newblock {\em Phys. Rev. D}, 84:054010, 2011.

\bibitem{Epelbaum:2006pt}
E.~Epelbaum and U.~G. Meissner.
\newblock {On the Renormalization of the One-Pion Exchange Potential and the
  Consistency of Weinberg`s Power Counting}.
\newblock {\em Few Body Syst.}, 54:2175--2190, 2013.


\end{thebibliography}
\end{document}